# Modelling and Simulation of the Propagation of P-SV Seismic Waves from Earthquakes: Application to Deep Earthquakes in Acre, Brazil

PEDRO HUAN MOREIRA[1], ANDINA ALAY LERMA[2], ELIANDRO RODRIGUES CIRILO[3], NEYVA MARIA LOPES ROMEIRO[4], WALDEMIR LIMA DOS SANTOS[5], PAULO LAERTE NATTI[6].

[1]Universidade Estadual de Londrina, Departamento de Matemática, Rodovia Celso Garcia Cid (PR 445), Km 380, 86057-970, Londrina - PR, Brazil.

ORCID http://orcid.org/0009-0005-6009-4365

[2]Universidade de Pernambuco, Escola Politécnica, Av. Engenheiro José Estelita, s/n, Santo Amaro, 50040-000, Recife, PE, Brazil.

ORCID http://orcid.org/0000-0001-7904-7773

[3]Universidade Estadual de Londrina, Departamento de Matemática, Rodovia Celso Garcia Cid (PR 445), Km 380, 86057-970, Londrina - PR, Brazil.

ORCID http://orcid.org/0000-0001-7530-1770

[4]Universidade Estadual de Londrina, Departamento de Matemática, Rodovia Celso Garcia Cid (PR 445), Km 380, 86057-970, Londrina - PR, Brazil.

ORCID http://orcid.org/0000-0002-3249-3490

[5]Universidade Estadual do Acre, Departamento de Geografia, Rodovia BR-364, Km 04, Bairro Distrito Industrial, 69.920-900, Rio Branco, AC, Brazil.

ORCID http://orcid.org/0000-0002-5306-5612

[6]Universidade Estadual de Londrina, Departamento de Matemática, Rodovia Celso Garcia Cid (PR 445), Km 380, 86057-970, Londrina - PR, Brazil,

email: plnatti@uel.br , ORCID http://orcid.org/0000-0002-5988-2621

**ABSTRACT**

Brazil is located in the central-eastern portion of the South American Plate, meaning that the country mostly experiences low-intensity seismic activity within its territory. However, some geological faults in this region have generated intense earthquakes. In this context, we intend to describe a recent earthquake of magnitude around 6.5 $M_b$ that occurred at a depth of approximately 600 km in the state of Acre, Brazil. In this work, we modeled the propagation of P-SV seismic waves using a two-dimensional system of partial differential equations (PDEs) in a two-dimensional vertical rectangular domain. The source is modeled by a Gaussian pulse function. The initial quiescence condition and Neumann boundary conditions are used. The PDE system is discretized by the finite difference method (FDM) and solved by the Gauss-Seidel method (GSM). The numerical simulations obtained describe the propagation of attenuated seismic waves in multiple geological layers, simulating intense and deep earthquakes in Acre. We used the propagation of perfect seismic waves to validate the model. The results include images of the simulations and theoretical seismograms simulating the vertical and horizontal displacement in the epicenter region and 200 km east and west of the epicenter.

## INTRODUCTION

Although the term earthquake is more commonly used for large-magnitude destructive events, while smaller ones are called tremors or seismic shocks, all are the result of the same geological process of slow accumulation and rapid release of energy, generating vibratory movements called seismic waves (Teixeira et al. 2009). The energy released by these events propagates in all directions in the form of seismic waves, which can be completely absorbed by the geological layers or reach the surface, manifesting as tremors that can cause damage. These events can be caused by geological faults, volcanic explosions, landslides, collisions between tectonic plates, or even human activity, such as the detonation of atomic bombs or the creation of reservoirs for hydroelectric power plants. When events of this type occur inside the Earth, seismic vibrations are generated that propagate in all directions in the form of waves, and it is these waves that cause damage and can be recorded by seismographs. Earthquakes are natural disasters that cannot be prevented. However, by studying these events and understanding them in detail, we can develop new technologies to mitigate the damage caused by this type of phenomenon. To this end, it is necessary to develop studies that allow the creation of seismic risk prediction strategies.

Brazil is considered a country with a low seismic risk. On average, earthquakes of magnitude ≥5 occur every five years in Brazil. Along the South Atlantic coast, from Rio Grande do Sul to Espírito Santo, an event of magnitude ≥5 occurs every 20 or 25 years. In contrast, in the Andean region, earthquakes of magnitude ≥5 occur on average twice a week. Most earthquakes are caused by geological faults in the crust of the South American plate, generating low-intensity earthquakes with hypocenters at depths between 10 km and 50 km, as observed in Figure 1 (Alcrudo, 2014).

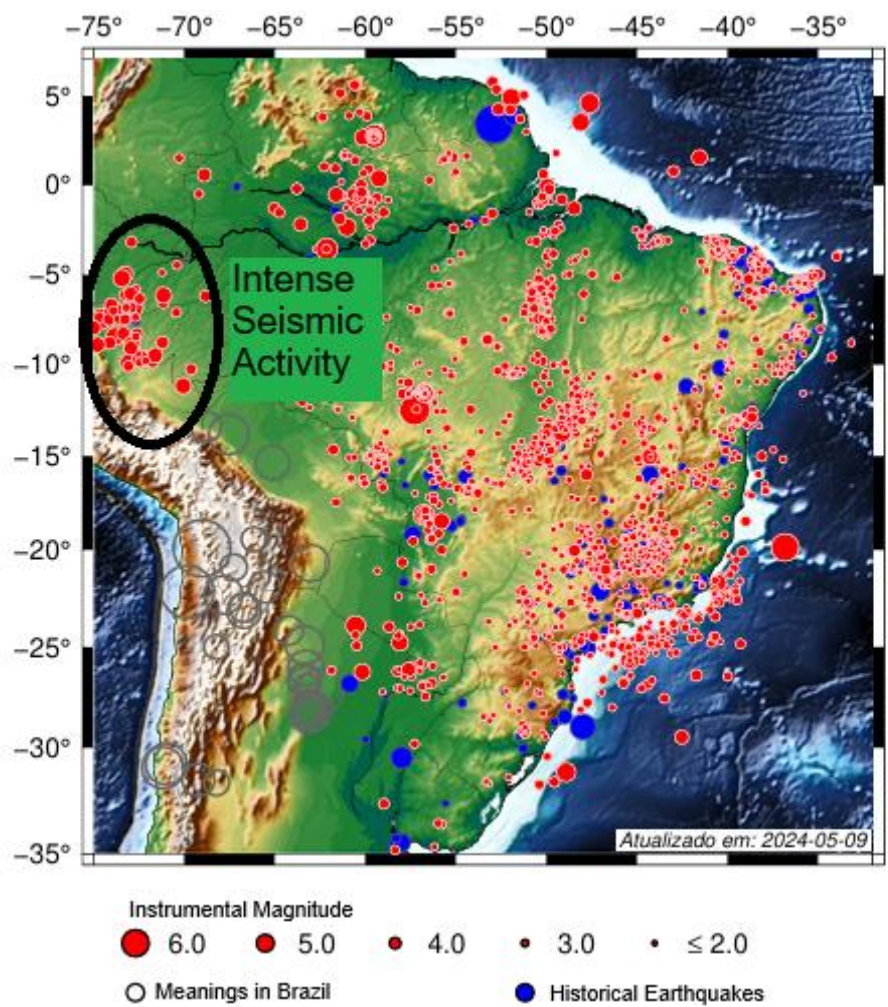

Figure 1: Known earthquakes of magnitude ≥ 2.8 in Brazil. Source: Adapted from Alcrudo (2014).

Despite relative stability, there are regions in Brazil where earthquakes of greater intensity occur. The region of Brazil with the highest seismic risk is the Northeast, due to its proximity to the Central Atlantic Fault, while the

North/Northwest of Brazil, due to its proximity to the Andes Mountains, experiences more intense and deeper earthquakes (Santos et al. 2010).

Specifically in the state of Acre, near the border with Peru, on 07-06-2022 at 9:55 pm (local time), an earthquake of magnitude 6.5 $M_b$ occurred at a depth of approximately 615 km and was recorded as the second largest earthquake in magnitude. More recently, the largest earthquake ever recorded in Brazil occurred on 20-01-2024 at 6:31 pm (local time), in the state of Acre with a magnitude 6.6 $M_b$. Table I presents the main seismic events that have occurred in Brazil since 2015, with a depth greater than 600 km.

As we can see in Table I and Figure 2, the region with the highest recorded deep seismic activity is the area near the municipalities of Tarauacá and Jordão, located in the state of Acre. The local sedimentary composition is mainly silt, clay, and sand. The region has an equatorial climate and a considerably high rainfall index, making the soil characteristically moist and clayey-sandy (Santos et al. 2019).

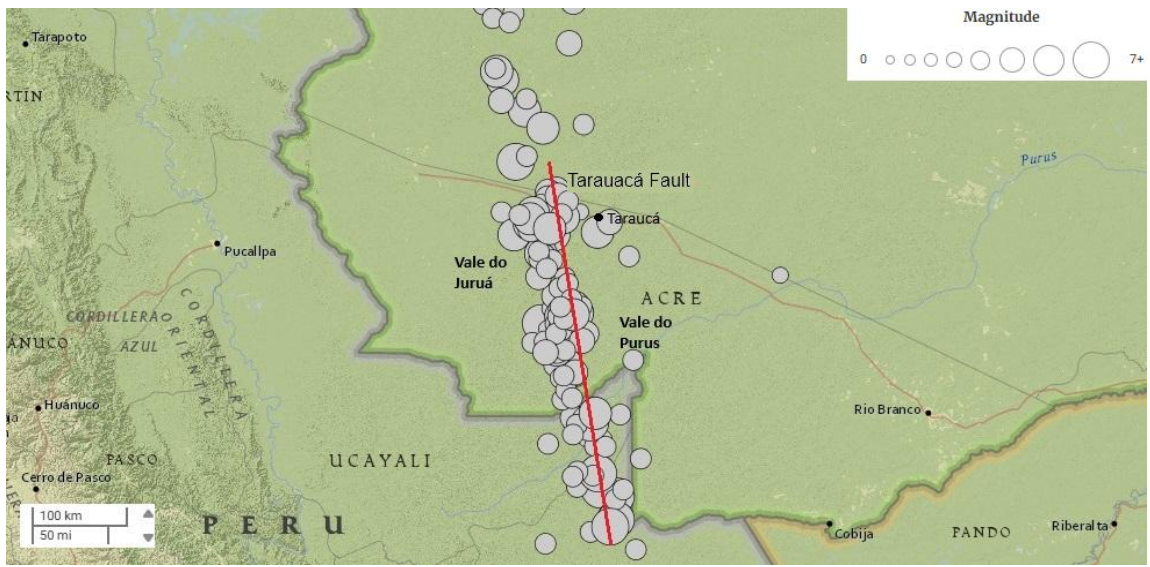

Figure 2: Tarauacá Fault - Earthquakes from 1950 onwards. Source: Adapted from Santos et al. (2019).

According to Santos et al. (2019), the frequency of earthquakes in the state of Acre has intensified in recent decades, especially since 2010, with a peak of 21 earthquakes recorded between 2013 and 2016. The authors concluded in their seismic studies that the state of Acre is divided into two plates contained within the South American plate, through a large geological fault called the "Tarauacá Fault", which runs North-South.

| DATA | HORA | PROFUNDIDADE | MAGNITUDE | ESCALA | REGIÃO |
|---|---|---|---|---|---|
| 2015-03-28 | 18:48:14 | 610,6 Km | 4,4 | mB | Jordão/AC |
| 2015-05-11 | 05:21:43 | 616,1 Km | 4,0 | mB | Jordão/AC |
| 2015-10-28 | 15:10:55 | 602,8 Km | 4,5 | mB | Santa Rosa do Purus/AC |
| 2015-11-24 | 23:05:00 | 615,0 Km | 4,1 | mB | Jordão/AC |
| 2015-11-24 | 21:55:20 | 617,0 Km | 4,2 | mB | Acre - Fronteira com Peru |
| 2015-11-25 | 16:47:55 | 660,0 Km | 4,7 | mB | Acre - Fronteira com Peru |
| 2015-11-25 | 07:26:22 | 651,3 Km | 4,9 | mB | Tarauacá/AC |
| 2015-11-26 | 22:52:54 | 627,0 Km | 5,1 | mB | Acre - Fronteira com Peru |
| 2015-11-26 | 04:01:23 | 605,9 Km | 4,8 | mB | Acre - Fronteira com Peru |
| 2015-11-26 | 03:45:18 | 604,9 Km | 6,2 | mB | Jordão/AC |
| 2015-11-29 | 08:41:31 | 647,4 Km | 4,1 | mB | Acre - Fronteira com Peru |
| 2016-01-03 | 09:08:34 | 631,0 Km | 4,7 | mB | Acre - Fronteira com Peru |
| 2016-08-30 | 06:34:28 | 631,2 Km | 3,9 | mB | Ipixuna/AM |
| 2016-08-31 | 05:42:18 | 616,7 Km | 3,7 | mB | Acre - Fronteira com Peru |
| 2016-12-18 | 11:30:11 | 640,0 Km | 6,2 | mB | Acre - Fronteira com Peru |
| 2018-08-26 | 16:58:25 | 635,9 Km | 4,6 | mB | Acre - Fronteira com Peru |
| 2021-10-12 | 01:34:34 | 603,0 Km | 4,3 | mB | Acre - Fronteira com Peru |
| 2022-06-07 | 22:53:28 | 611,7 Km | 5,1 | mB | Acre - Fronteira com Peru |
| **2022-06-07** | **21:55:45** | **615,0 Km** | **6,5** | **mB** | **Acre - Fronteira com Peru** |
| 2022-07-24 | 05:17:08 | 613,0 Km | 4,4 | mB | AM/AC |
| 2022-11-11 | 09:34:46 | 643,0 Km | 5,0 | mB | Reg. de Tarauacá/AC |
| 2023-01-17 | 06:59:25 | 639,0 Km | 4,4 | mB | Reg. de Tarauacá/AC |
| **2024-01-20** | **18:31:07** | **638,1 Km** | **6,6** | **mB** | **Acre - Fronteira com Peru** |
| 2024-01-28 | 06:38:59 | 651,7 Km | 6,2 | mB | Acre - Fronteira com Peru |

Table I: Recent deepest seismic events in Brazilian territory. Source: Adapted from USP Seismology Center (USP, 2024)

Despite the frequency of intense earthquakes in the Tarauacá region, the damage caused by these tremors is small or nonexistent, as they occur at great depths and are strongly attenuated by the geological layers. Even so, the seismic risk in this region is not negligible, and further studies are needed to better understand these events. This is the focus of this study.

## MATERIALS AND METHODS

### Types of seismic waves and seismic scale

Primary (P) waves generate expansion/compression deformations in the medium and are analogous to sound waves propagating through the air, being fast and the first to reach the seismograph (Press et al. 2004). Due to these waves, oscillations in the medium occur parallel to the direction of wave propagation, causing successive compressions and expansions that shape the movement of the medium where the P waves pass.

Secondary waves (SV and SH), also called shear waves, are slower and therefore take longer to reach the seismograph. Although their speed is lower, their destructive power is greater, as secondary waves generate oscillations in the medium in a direction perpendicular to the direction of propagation of these waves. In the horizontal direction, we have SH waves and, in the vertical direction, SV waves.

There are also surface waves, which arise from interactions between P and S waves close to the surface, generating Love and Rayleigh waves. These waves are the most destructive, as they have the greatest potential to shake structures and cause damage.

In this work, to measure the magnitude of simulated earthquakes, we will use the $M_b$ scale (body wave magnitude), defined by

$$M_b = \log\left(\frac{A}{T}\right) + B , \qquad (1)$$

where $A$ is the amplitude of the primary wave P, in micro metros ($\mu m$), $T$ is the period of oscillation of P wave or S wave, in seconds, and $B$ is a correction distance function from the epicenter determined empirically by the Gutenberg-Richter formula (Gutenberg & Richter, 1956), which was later revised by Aki & Richards et al. (2009).

**Mathematical modeling of P-SV seismic wave propagation with attenuation**

Consider a Galilean reference frame. To mathematically model the interaction of seismic waves in a solid medium, let us consider a control volume and the forces acting on its faces, as shown in Figure 3.

We decompose the resultant stress and strain force acting on this surface into three spatial components and denote each of these components by $\tau_{ij}$, where $i$ represents the normal direction of the surface, and $j$ represents the direction in which the force is acting. We are interested in how these forces vary in each direction and on each surface, since these combined variations are what cause the stresses and strains in the control volume. Consider the force variations on a surface with normal direction $i$ given by

$$F_i = \left(\frac{\partial \tau_{ix}}{\partial x} + \frac{\partial \tau_{iy}}{\partial y} + \frac{\partial \tau_{iz}}{\partial z}\right) \delta x \delta y \delta z \ . \tag{2}$$

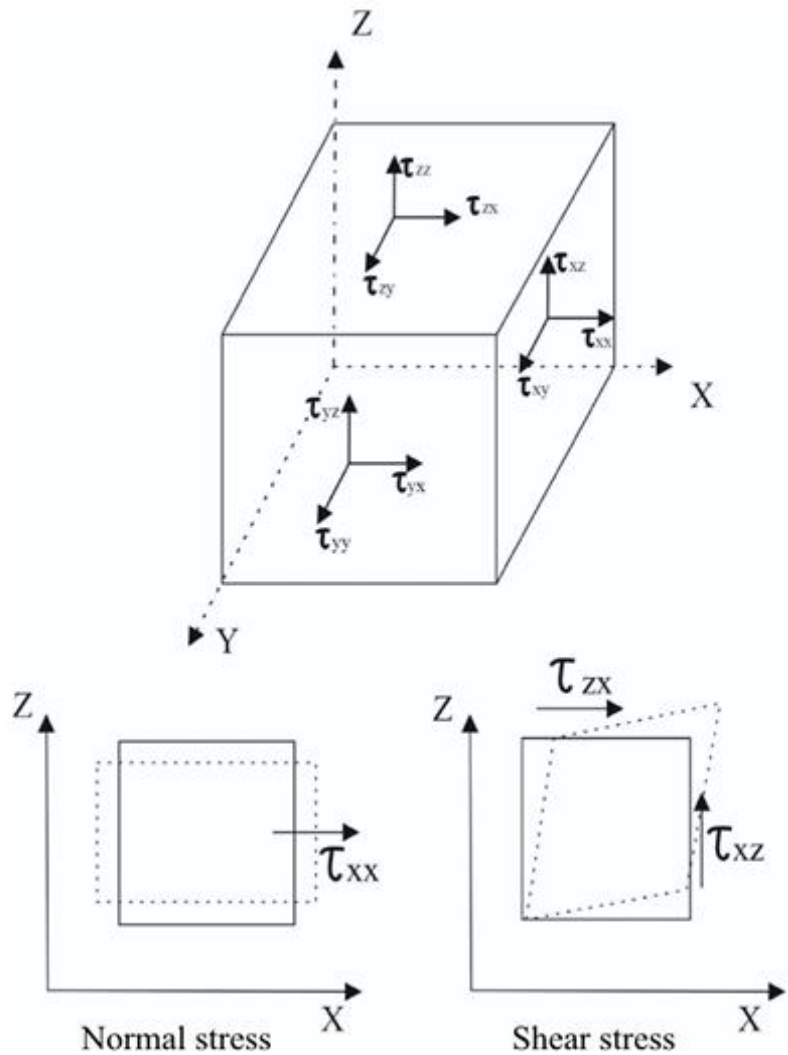

Figure 3: Forces acting on the faces of a control volume of an elastic medium. Source: Adapted from Alay (2021).

We are also interested in considering an attenuating force acting on the control volume, causing the seismic wave to lose energy as it propagates. Thus, we add an attenuation force to the model (2),

$$f_i = -\gamma \, \frac{\partial u_i}{\partial t} \, \delta x \delta y \delta z \ , \tag{3}$$

where $u_i$ is the displacement in the direction $i$, and $\gamma$ is the attenuation coefficient. From Newton's second law in direction $i$, we can obtain the acceleration on this face of control volume through the second derivative of the displacement,

$$\rho \frac{\partial^2 u_i}{\partial t^2} = \frac{\partial \tau_{ix}}{\partial x} + \frac{\partial \tau_{iy}}{\partial y} + \frac{\partial \tau_{iz}}{\partial z} - \gamma \frac{\partial u_i}{\partial t}, \qquad (4)$$

where $\rho = m/(\delta x \delta y z \delta z)$ is the density of the medium.

For the case of isotropic means, the Hooke's law gets simpler and allows us to describe the strain stresses $\tau_{ij}$ with only two coefficients, the Lamé coefficients $\mu$ and $\lambda$. The coefficient $\mu$, known as the shear modulus, measures resistance to shape change, while the coefficient $\lambda$, known as the Lame's first parameter, relates to resistance to volume change (bulk compression) (Tago et al., 2012). The mathematical expression is given by

$$\tau_{ij} = \lambda \delta_{ij} \left( \frac{\partial u_x}{\partial x} + \frac{\partial u_y}{\partial y} + \frac{\partial u_z}{\partial z} \right) + \mu \left( \frac{\partial u_j}{\partial i} + \frac{\partial u_i}{\partial j} \right), \qquad (5)$$

where $\delta_{ij}$ is the Kronecker delta.

For our two-dimensional vertical study, let us consider only the displacements in the $x$ and $z$ axes, i.e., $u = (u_x, 0, u_z)$, simulating P and SV waves. Then, combining equations (4) and (6) in the two-dimensional vertical case, we obtain

$$\rho \frac{\partial^2 u_x}{\partial t^2} = (\lambda + 2\mu) \frac{\partial^2 u_x}{\partial x^2} + \mu \frac{\partial^2 u_x}{\partial z^2} + (\lambda + \mu) \frac{\partial^2 u_z}{\partial z \partial x} - \gamma \frac{\partial u_x}{\partial t} \qquad (4)$$

$$\rho \frac{\partial^2 u_z}{\partial t^2} = (\lambda + 2\mu) \frac{\partial^2 u_z}{\partial z^2} + \mu \frac{\partial^2 u_z}{\partial x^2} + (\lambda + \mu) \frac{\partial^2 u_x}{\partial z \partial x} - \gamma \frac{\partial u_z}{\partial t}. \qquad (5)$$

This system of elastodynamic equations is frequently used for seismic wave simulations. Vireaux (1986), Levander (1988), Tago et al. (2012), Alay (2021) and others, used the same model to develop their work in seismic wave studies. Tago et al. (2012) and Jacobs et al. (1974) show that the speed of P and S waves can be functions of the Lamé parameters and density,

$$v_P = \sqrt{\frac{\lambda + 2\mu}{\rho}} \quad \text{and} \quad v_S = \sqrt{\frac{\mu}{\rho}}. \qquad (6)$$

So, rewriting the equation in terms of speeds $v_P^2$ and $v_S^2$, we obtain

$$\frac{\partial^2 u_x}{\partial t^2} = v_P^2 \left(\frac{\partial^2 u_x}{\partial x^2} + \frac{\partial^2 u_z}{\partial z \partial x}\right) + v_S^2 \left(\frac{\partial^2 u_x}{\partial z^2} - \frac{\partial^2 u_z}{\partial z \partial x}\right) - \frac{\gamma}{\rho}\frac{\partial u_x}{\partial t} \qquad (7)$$

$$\frac{\partial^2 u_z}{\partial t^2} = v_P^2 \left(\frac{\partial^2 u_z}{\partial z^2} + \frac{\partial^2 u_x}{\partial z \partial x}\right) + v_S^2 \left(\frac{\partial^2 u_z}{\partial x^2} - \frac{\partial^2 u_x}{\partial z \partial x}\right) - \frac{\gamma}{\rho}\frac{\partial u_z}{\partial t}, \qquad (8)$$

the PDE system that we will use in this work to model and simulate the propagation of P-SV seismic waves.

**Initial and boundary conditions**

In this work we use a rectangular domain to simulate the propagation of P-SV seismic waves. For the initial condition, the medium is assumed to be in a steady state initially, so that there are no displacements $u_x$ and $u_z$, in directions $x$ and $z$ respectively, at $t = 0$. Therefore, in the domain $\Omega$ of the studied problem, we impose that.

$$u_x(x,z,t) = 0 \quad \text{and} \quad u_z(x,z,t) = 0 . \qquad (9)$$

For the boundary conditions in the domain, shown in Figure 4, we impose Neumann-type conditions for all boundaries. Figure 4 can be understood as a vertical domain within the Earth. Segments B1, B3, and B4 are domain boundaries, inside the Earth, while segment B2 represents Earth's surface. The Neumann conditions for equation (7) are given by

$$\frac{\partial u_x(B1)}{\partial n} = \frac{\partial u_x(B2)}{\partial n} = \frac{\partial u_x(B3)}{\partial n} = \frac{\partial u_x(B4)}{\partial n} = 0 , \qquad (10)$$

while the Neumann conditions for the equation (8) are

$$\frac{\partial u_z(B1)}{\partial n} = \frac{\partial u_z(B2)}{\partial n} = \frac{\partial u_z(B3)}{\partial n} = \frac{\partial u_z(B4)}{\partial n} = 0 . \qquad (11)$$

A boundary attenuation condition is applied. In Cerjan et al. (1985) and Bai et al. (2013), exponential attenuations in the displacements $u_x$ and $u_z$ of the wave are applied. In this work, we applied attenuation $\sigma_x(x)$ and $\sigma_y(x)$ using an algebraic function in the velocities $v_P$ and $v_S$, as the results were more efficient,

$$\sigma_x(x) = \left(\frac{x-N}{N}\right)^{\frac{3}{2}} \quad \text{and} \quad \sigma_z(z) = \left(\frac{z-N}{N}\right)^{\frac{3}{2}}, \qquad (12)$$

where $N$ is the size of the attenuating layer, $x$ and $z$ are positions of the wavefront in the domain.

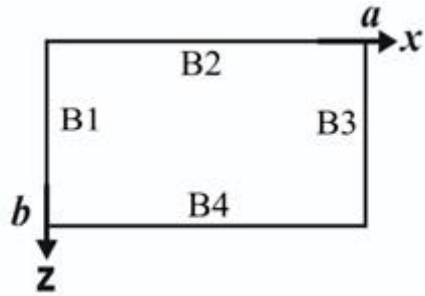

Figure 4: Geometric representation of boundaries. Source: Alay (2021)

Regarding the initial energy of the earthquake, Tago et al. (2012) use the Dirac delta function as a point source, at hypocenter $(x_i, z_i)$. However, it is more common to use the Gaussian function, as proposed in the pioneering work of Alterman and Karal (1968). In this work, we use the Gaussian function (13) to simulate the earthquake source,

$$E(x,z,t) = a \exp(-ct^2) \cdot \exp(-\kappa(x^2 + z^2)), \qquad (13)$$

where the amplitude $a$ is measured in $N\frac{m^2}{Kg}$; $\kappa$ is the spatial decay parameter, measured in $m^{-2}$; and $c$ is the time decay parameter, measured in $Hz^{-2}$.

**Numerical Model**

To solve the system of elastodynamic equations (7) and (8), we used the finite difference method, with a second order approximation scheme in time and space. For spatial derivatives we used a centered finite difference scheme, while for derivatives in time we used a regressive finite difference scheme. For the boundary conditions, the spatial derivatives were discretized in a regressive derivative scheme.

To simulate the earthquakes that have occurred in the state of Acre, Brazil, at depths on the order of 600 km, we defined the problem domain as $[x_0, x_f]$ = [-600,600] and $[z_0, z_f]$ = [-600, 600], with the center of the domain being the hypocenter of the earthquake. It is necessary to discretize the domain so that we can treat it computationally. To do this, it is necessary to define the elements $\Delta_x$ and $\Delta_z$, which are the spatial increments, as

$$\Delta_x = \frac{x_f - x_0}{n_x} \quad \text{and} \quad \Delta_z = \frac{z_f - z_0}{n_z}. \tag{14}$$

The quantities $n_x$ and $n_z$ in (14) are integers that determine the number of points in each direction of the discretized mesh. The higher $n_x$ and $n_z$, the better the spatial approximation, but the computational cost increases. In our simulations, we did $\Delta_t = \frac{\Delta_x}{2} = \frac{\Delta_z}{2}$ for convergence purposes.

To obtain the numerical solution of the P-SV seismic wave system, a numerical scheme given by the FDM is used (Oliveira et al., 2019; Dall'Agnol et al., 2019). This numerical modeling proposed gave an approximation to the solution of the problem in time $(l+1)$ and in the position $(i,k)$ of the computational mesh, namely,

$$\frac{\partial^2 u_x}{\partial t^2}\Big|_{i,k}^{l+1} = v_P^2 \left(\frac{\partial^2 u_x}{\partial x^2}\Big|_{i,k}^{l+1} + \frac{\partial^2 u_z}{\partial z \partial x}\Big|_{i,k}^{l+1}\right) + v_S^2 \left(\frac{\partial^2 u_x}{\partial z^2}\Big|_{i,k}^{l+1} - \frac{\partial^2 u_z}{\partial z \partial x}\Big|_{i,k}^{l+1}\right) - \frac{\gamma}{\rho} \frac{\partial u_x}{\partial t}\Big|_{i,k}^{l+1} + E\Big|_{i,k}^{l+1} \tag{15}$$

$$\frac{\partial^2 u_z}{\partial t^2}\Big|_{i,k}^{l+1} = v_P^2 \left(\frac{\partial^2 u_z}{\partial z^2}\Big|_{i,k}^{l+1} + \frac{\partial^2 u_x}{\partial z \partial x}\Big|_{i,k}^{l+1}\right) + v_S^2 \left(\frac{\partial^2 u_z}{\partial x^2}\Big|_{i,k}^{l+1} - \frac{\partial^2 u_x}{\partial z \partial x}\Big|_{i,k}^{l+1}\right) - \frac{\gamma}{\rho} \frac{\partial u_z}{\partial t}\Big|_{i,k}^{l+1} + E\Big|_{i,k}^{l+1}. \tag{16}$$

To solve our system of equations (15-16), we can rewrite our discretized model as

$$u_x\big|_{i,k}^{l+1} = \frac{1}{A_{Px}}\left[A_{Ex} u_x\big|_{i+1,k}^{l+1} + A_{Wx} u_x\big|_{i-1,k}^{l+1} + A_{Nx} u_x\big|_{i,k+1}^{l+1} + A_{Sx} u_x\big|_{i,k-1}^{l+1} + b1 + ACu_z + E\big|_{i,k}^{l+1}\right] \tag{17}$$

$$u_z\big|_{i,k}^{l+1} = \frac{1}{A_{Pz}}\left[A_{Ez} u_z\big|_{i+1,k}^{l+1} + A_{Wz} u_z\big|_{i-1,k}^{l+1} + A_{Nz} u_z\big|_{i,k+1}^{l+1} + A_{Sz} u_z\big|_{i,k-1}^{l+1} + b2 + ACu_x + E\big|_{i,k}^{l+1}\right], \tag{18}$$

where

$$A_{Px} = \frac{2}{\Delta t^2} + \frac{2v_P^2}{\Delta x^2} + \frac{2v_S^2}{\Delta z^2} + \frac{3\gamma}{2\rho \Delta t} \tag{19}$$

$$A_{Ex} = A_{wx} = \frac{v_p^2}{\Delta x^2} \tag{20}$$

$$A_{Nx} = A_{sx} = \frac{v_s^2}{\Delta z^2} \tag{21}$$

$$b_x = \left(\frac{5}{\Delta t^2} + \frac{4\gamma}{2\rho\Delta t}\right) u_{x_{i,k}}^l - \left(\frac{4}{\Delta t^2} + \frac{\gamma}{2\rho\Delta t}\right) u_{x_{i,k}}^{l-1} + \frac{u_{x_{i,k}}^{l-2}}{\Delta t^2} \tag{22}$$

$$A_{Cx} u_z^{l+1} = (v_p^2 - v_s^2) \left(\frac{u_{z_{i+1,k+1}}^{l+1} - u_{z_{i+1,k-1}}^{l+1} + u_{z_{i-1,k-1}}^{l+1} - u_{z_{i-1,k+1}}^{l+1}}{4\Delta x \Delta z}\right) \tag{23}$$

and

$$A_{Pz} = \frac{2}{\Delta t^2} + \frac{2v_p^2}{\Delta z^2} + \frac{2v_s^2}{\Delta x^2} + \frac{3\gamma}{2\rho\Delta t} \tag{24}$$

$$A_{Ez} = A_{wz} = \frac{v_p^2}{\Delta z^2} \tag{25}$$

$$A_{Nz} = A_{sz} = \frac{v_s^2}{\Delta x^2} \tag{26}$$

$$b_z = \left(\frac{5}{\Delta t^2} + \frac{4\gamma}{2\rho\Delta t}\right) u_{z_{i,k}}^l - \left(\frac{4}{\Delta t^2} + \frac{\gamma}{2\rho\Delta t}\right) u_{z_{i,k}}^{l-1} + \frac{u_{z_{i,k}}^{l-2}}{\Delta t^2} \tag{27}$$

$$A_{Cz} u_x^{l+1} = (v_p^2 - v_s^2) \left(\frac{u_{x_{i+1,k+1}}^{l+1} - u_{x_{i+1,k-1}}^{l+1} + u_{x_{i-1,k-1}}^{l+1} - u_{x_{i-1,k+1}}^{l+1}}{4\Delta x \Delta z}\right), \tag{28}$$

**Discretization of initial and boundary conditions**

The boundary conditions obtained using the finite difference method used a second order back-difference scheme (Cirilo et al., 2018), such that the discretized boundary conditions are expressed as

$$\frac{\partial u_x}{\partial x}(Bn)|_{i,k}^{l+1} = \frac{3u_{x_{i,k}}^{l+1} - 4u_{x_{i-1,k}}^{l+1} + u_{x_{i-2,k}}^{l+1}}{2\Delta x} = 0 \Rightarrow u_{x_{i,k}}^{l+1} = \frac{4u_{x_{i-1,k}}^{l+1} - u_{x_{i-2,k}}^{l+1}}{3} \tag{29}$$

$$\frac{\partial u_z}{\partial x}(Bn)|_{i,k}^{l+1} = \frac{3u_{z_{i,k}}^{l+1} - 4u_{z_{i-1,k}}^{l+1} + u_{z_{i-2,k}}^{l+1}}{2\Delta x} = 0 \Rightarrow u_{z_{i,k}}^{l+1} = \frac{4u_{z_{i-1,k}}^{l+1} - u_{z_{i-2,k}}^{l+1}}{3}, \tag{30}$$

$$\frac{\partial u_x}{\partial z}(Bn)|_{i,k}^{l+1} = \frac{3u_{x_{i,k}}^{l+1} - 4u_{x_{i-1,k}}^{l+1} + u_{x_{i-2,k}}^{l+1}}{2\Delta z} = 0 \Rightarrow u_{x_{i,k}}^{l+1} = \frac{4u_{x_{i-1,k}}^{l+1} - u_{x_{i-2,k}}^{l+1}}{3} \tag{31}$$

$$\frac{\partial u_z}{\partial z}(Bn)|_{i,k}^{l+1} = \frac{3u_{z_{i,k}}^{l+1} - 4u_{z_{i-1,k}}^{l+1} + u_{z_{i-2,k}}^{l+1}}{2\Delta z} = 0 \Rightarrow u_{z_{i,k}}^{l+1} = \frac{4u_{z_{i-1,k}}^{l+1} - u_{z_{i-2,k}}^{l+1}}{3}. \tag{32}$$

### Structure of computational algorithms

The algorithm used to solve the system of equations and obtain the simulation results was written in Python, using the Numpy and Matplotlib scientific computing packages. The solution of the linear system, resulting from the discretization, was obtained using the Gauss-Seidel method (Cirilo et al., 2019).

## RESULTS AND DISCUSSION

### Validation of numerical model and computational implementation

The system of equations (7) and (8) describes the propagation of seismic waves in a two-dimensional domain. Note that it is possible to validate the computational implementation (numerical model) by defining appropriate values for the parameters of the mathematical model. When we set $\mu = 1$, $\lambda = -1$, and $\gamma = 0$, equations (4-5) become decoupled PDEs for perfect waves (Alay 2021). Taking $\mu = 1$ and $\lambda = -1$ in definitions in (6), we have that $v_P = v_S = v$, then replacing in the system (7-8), we obtain PDEs of perfect waves, subject to external forces $E(x, z, t)$,

$$\frac{\partial^2 u_x}{\partial t^2} = v^2 \left( \frac{\partial^2 u_x}{\partial x^2} + \frac{\partial^2 u_x}{\partial z^2} \right) + E \tag{33}$$

$$\frac{\partial^2 u_z}{\partial t^2} = v^2 \left( \frac{\partial^2 u_z}{\partial x^2} + \frac{\partial^2 u_z}{\partial z^2} \right) + E \ . \tag{34}$$

Here the P and SV waves travel at the same speed, forming a system of independent perfect wave equations. In a perfect wave, the pressure and shear forces are equal. In this case, the simulation results are illustrated in figures 5 and 6, where the following values were adopted for the model parameters: $v_P = v_S = 7.76 \frac{Km}{s}$, $\gamma = 0 \frac{Km}{m.s}$, $a = 1.0 \frac{N.m^2}{Kg}$, $c = 2.0 \ Hz$, $\kappa = 1 \ m^{-2}$, $n_x = 300$, $n_z = 300$, $\Delta x = 4.0 \ km$, $\Delta z = 4.0 \ km$, and $\Delta t = 0.2s$.

The numerical simulations presented in Figures 5 and 6 describe, as expected, the propagation of a perfect two-dimensional wave. The reflections observed in Figure (6), at the boundary $B2$ of the computational domain, are consequences of abrupt changes in the wave propagation velocities (ground-air), while at the boundary $B1, B3$ and $B4$ of the computational domain no reflections occur, as desired, because the waves must pass through such boundaries. Note that we applied attenuating edge conditions in the vicinity of the boundaries $B1, B3$ and $B4$ to avoid unwanted reflections. These results are important because they behave as a validation test of the mathematical model, the numerical discretization, and the computational implementation. The region delimited externally by the green lines and by the domain boundaries corresponds to the points of the computational mesh where the attenuation condition (12) acts. The red dotted lines delineate different geological layers.

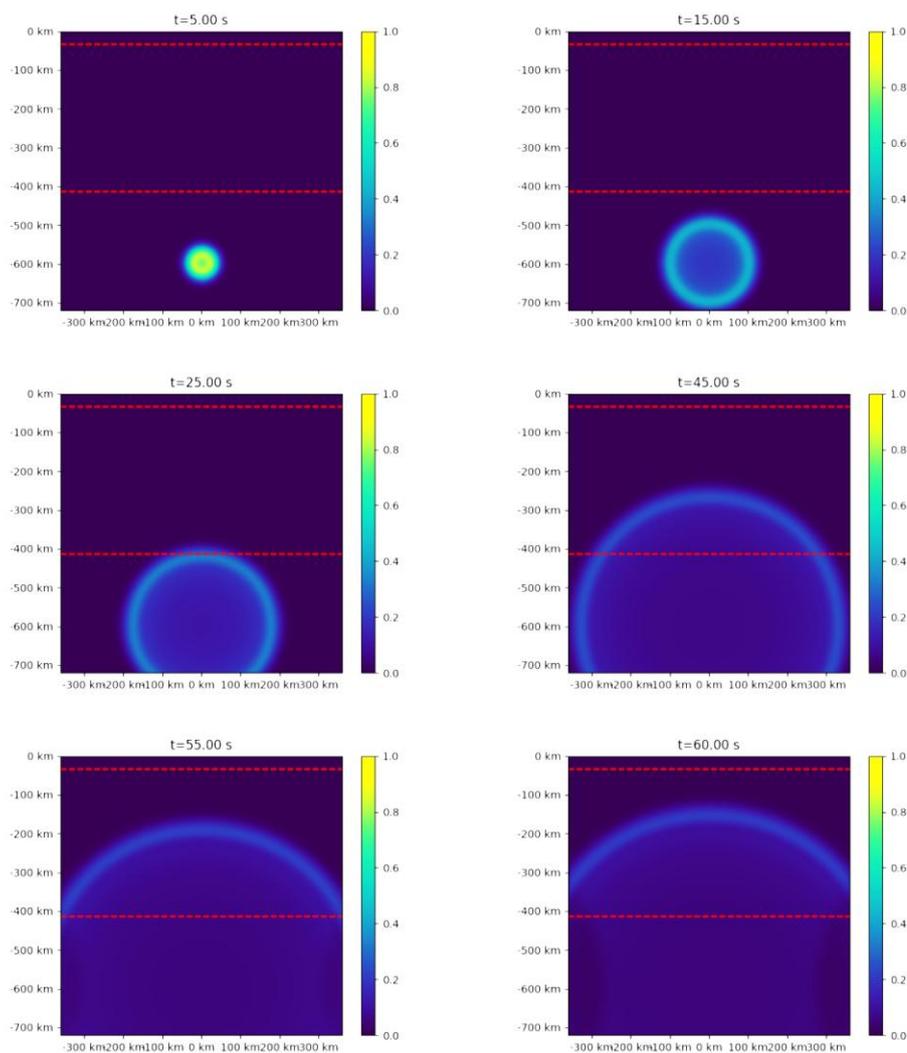

Figure 5: Simulation of the PDEs of uncoupled perfect waves, before the waves reach the upper boundary of the computational domain. Source: Author.

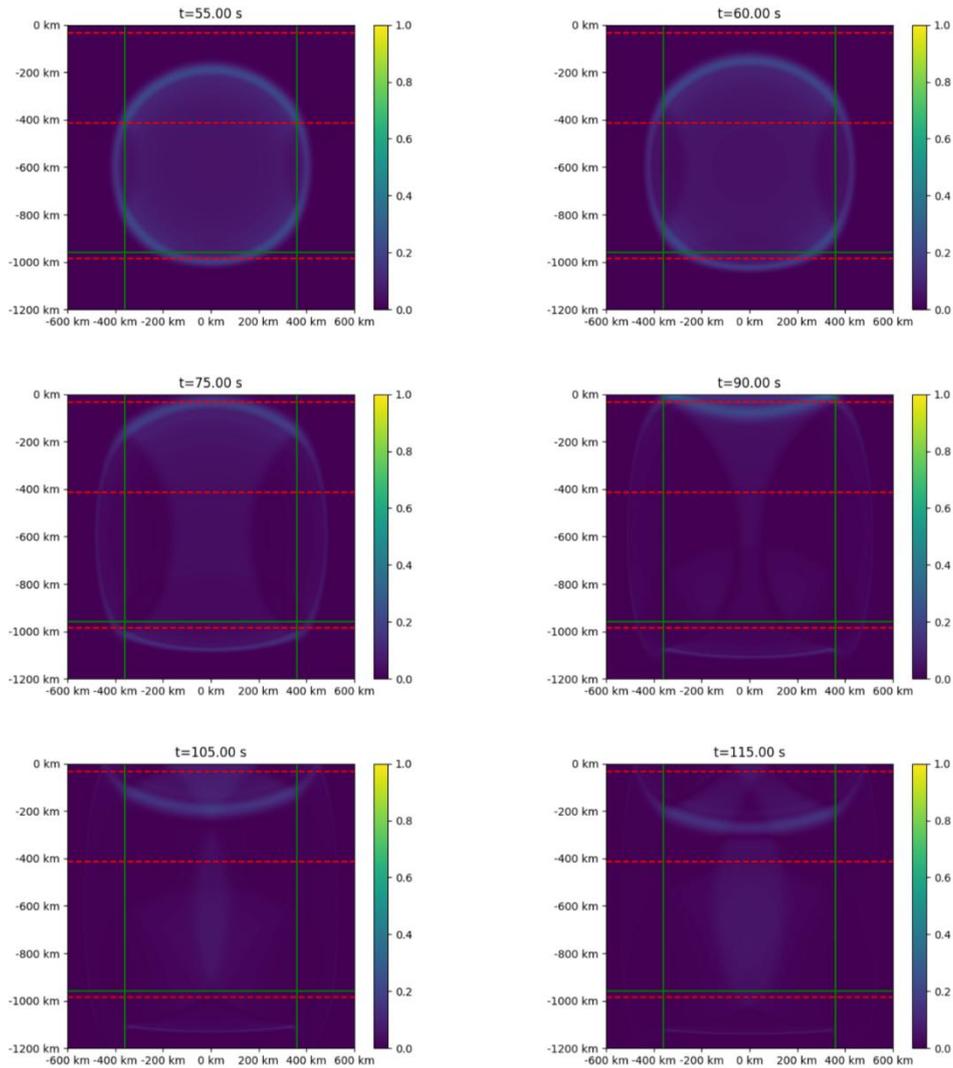

Figure 6: Simulation of the PDEs of uncoupled perfect waves at times before and after the waves reach the upper boundary of the computational domain. Source: Author.

In these simulations of perfect seismic waves, it was assumed that the geological layers had the same properties in order to avoid reflections. The next step is to obtain the attenuated propagation of P-SV waves, passing through different geological layers to the surface, to simulate earthquakes that occur in Acre.

**Dynamics of P-SV waves with attenuation in multiple geological layers**

Now, we present our simulations of recent earthquakes that occurred in the state of Acre, at a depth of approximately 600 km. This type of earthquake is uncommon in Brazilian territory, due to its depth and intensity.

Seismic waves undergo changes in speed depending on the region through which they propagate. The depth of 600 km is in a region of the upper mantle. In this context, we must consider that $v_P$ and $v_S$ are no longer constants, but functions that depend on the depth at which the seismic wave propagates, that is, $v_P(z)$ and $v_S(z)$. According to Jacobs et al. (1974), the main divisions within the Earth are represented in Table II, which describes the Earth's layers according to their physical properties.

| Region | | Maximum depth (Km) | Earth Fraction | Features |
|---|---|---|---|---|
| Crust | A | 0 | - | Surface |
| Crust / Mantle | B | 33 | 0.0155 | Very heterogeneous conditions |
| Mantle | C | 413 | 0.1667 | Probably homogeneous |
| Mantle | D | 984 | 0.2131 | Transition region |
| Mantle / Corner | E | 2898 | 0.4428 | Probably homogeneous |
| Corner | F | 4982 | 0.1516 | Homogeneous fluid |
| Corner | G | 5121 | 0.0028 | Transition layer |
| Corner | G | 6371 | 0.0075 | Inner core |

Table II: Depths and descriptions of the Earth's inner layers. Source: Adapted from Jacobs et al. (1974)

Regarding the attenuation coefficient $\gamma$ and density $\rho$ in equations (7-8), since the Earth's crust and mantle have different materials in their composition, it is not possible to determine these coefficients precisely. Teixeira et al. (2009) provide the average density $\rho$ of the crust and of the mantle, so in the crust $\rho = 2.7 \frac{g}{cm^3}$ and in the mantle $\rho = 4.5 \frac{g}{cm^3}$. For the clayey soil of Acre, Marcolin & Klein (2011) estimate an average density of $\rho = 1.77 \frac{g}{cm^3}$. Therefore, the parameters used in these simulations are given in Table III.

| | | Maximum depth (Km) | γ parameter | Density ρ g/cm³ | Atenuattion γ/ρ (Kg/m s) |
|---|---|---|---|---|---|
| Crust | A | 5 | 0,05 | 1,77 | 0,0282 |
| Crust / Mantle | B | 33 | 0,05 | 2,70 | 0,0185 |
| Mantle | C | 413 | 0,05 | 4,50 | 0,0111 |
| Mantle | D | 984 | 0,05 | 4,50 | 0,0111 |
| Mantle / Corner | E | 2898 | 0,05 | 4,50 | 0,0111 |

Table III: Values of the seismic wave attenuation coefficient and density as a function of the depth of geological layers. Source: Adapted from Moreira (2024)

With regard to velocities in equations (7-8), Jacobs et al. (1974) calibrated the P and S wave velocities, considering the results in Tables II and III, as a function of the depths of the Earth's layers. Table IV presents these adjusted values.

| Region | | Maximum depth (Km) | $\frac{\gamma}{\rho}(z)$ Kg/m.s | $v_P(z)$ Km/s | $v_S(z)$ Km/s |
|---|---|---|---|---|---|
| Crust | A | 0 | 0,0282 | 1,75 | 4,36 |
| Crust / Mantle | B | 33 | 0,0185 | 7,76 | 4,36 |
| Mantle | C | 413 | 0,0111 | 8,97 | 4,96 |
| Mantle | D | 984 | 0,0111 | 10,41 | 5,77 |
| Mantle / Corner | E | 2898 | 0,0111 | 13,64 | 7,30 |

Table IV: Parameters used in simulations of multiple geological layers. Source: Adapted from Moreira (2024)

Considering the data described in Table IV, the results of simulations of equations of propagation of P-SV seismic waves are presented in Figures 7 and 8, where the following values were adopted for the model parameters: $a = 1.0 \frac{N.m^2}{Kg}$, $c = 2.0\ Hz$, $\kappa = 1\ m^{-2}$, $n_x = 300$, $n_z = 300$, $\Delta x = 4.0\ km$, $\Delta z = 4.0\ km$, and $\Delta t = 0.2s$.

Unlike the numerical simulation of perfect seismic wave propagation presented in Figures (5-6), where P and SV waves had the same velocities and we see only one wavefront, now in Figures (7-8) we see two wavefronts, which propagate with velocities $v_P(z)$ and $v_S(z)$, with $v_P(z) > v_S(z)$ except in the crust.

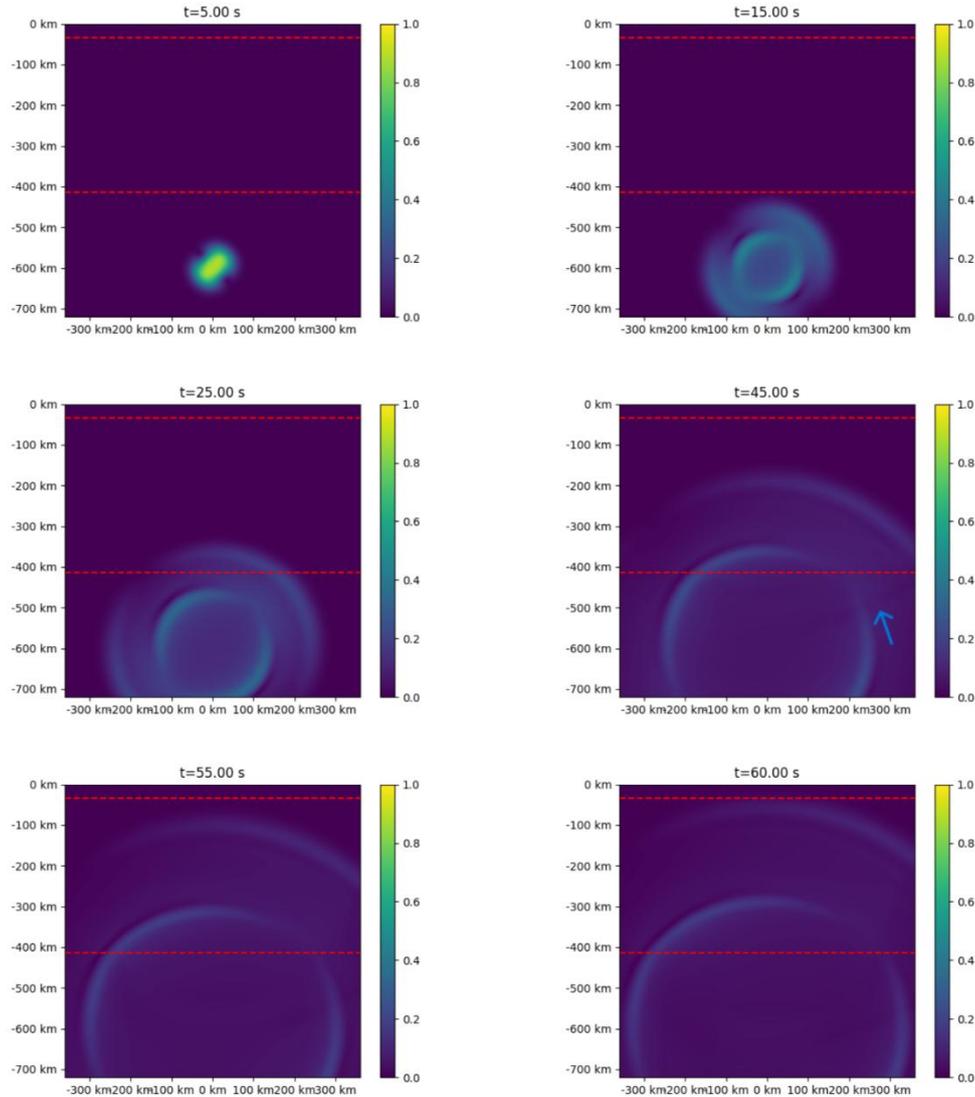

Figure 7: Simulation of PDEs of coupled P-SV waves, with different attenuations and velocities in the geological layers, before reaching the edges of the computational domain. Source: Author.

In Figure 7, at $t = 45.00$ seconds and $t = 55:00$ seconds, reflections can be observed near the 413 km, on the dividing line between layers D and C of Table IV. These reflections and refractions occur in the transition regions between layers, due to differences in the physical and chemical properties of the layers (mantle/upper mantle). These changes in physical and chemical properties are also responsible for changes in the speed of seismic waves.

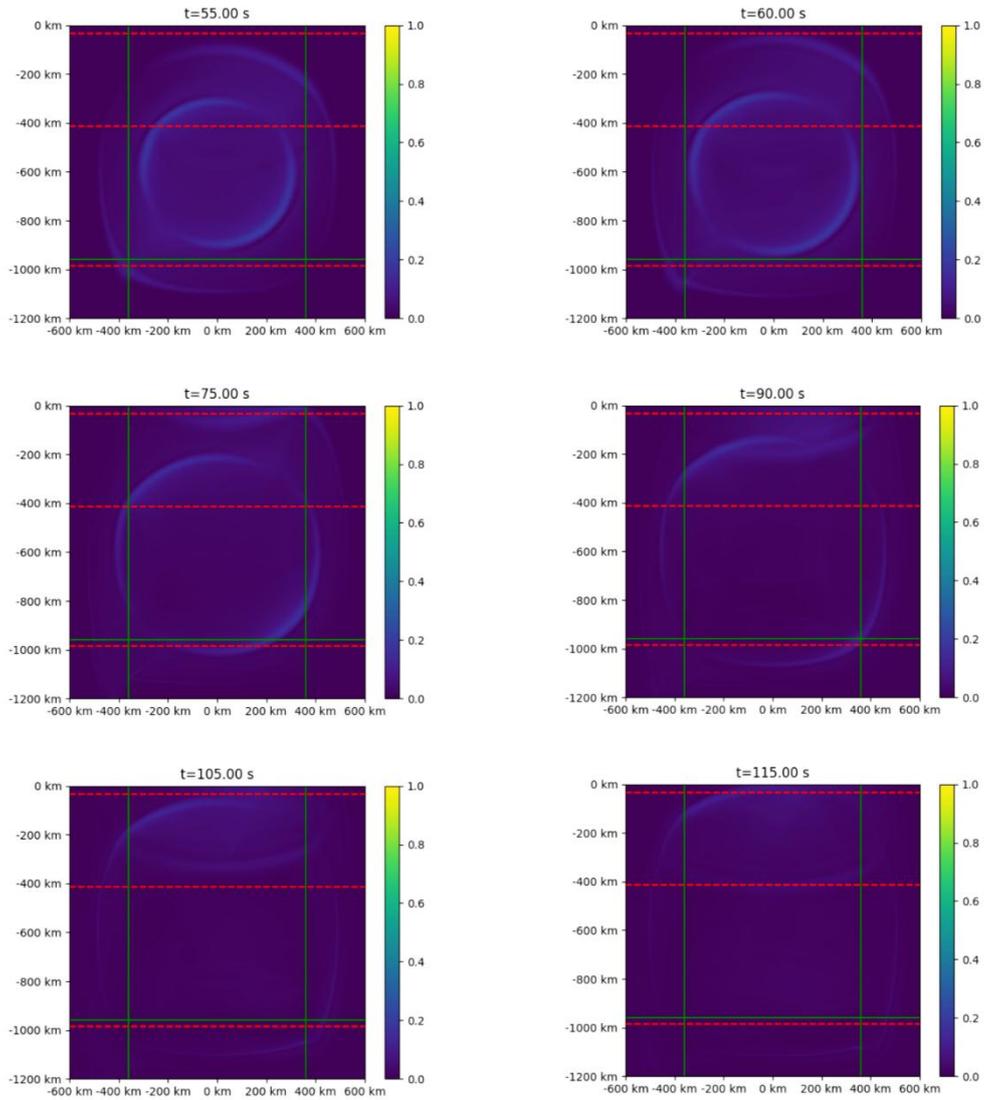

Figure 8: Simulation of PDEs of coupled P-SV waves, with different attenuations and velocities in the geological layers, after reaching the edges of the computational domain. Source: Author.

**Seismograms**

Another important result obtained from these numerical simulations was the theoretical seismograms. The seismograms were computationally calculated to show the amplitudes of the seismic waves, over a 300-second interval, at the epicenter and 200 km east and west of the epicenter. The results obtained are shown in Figure 9.

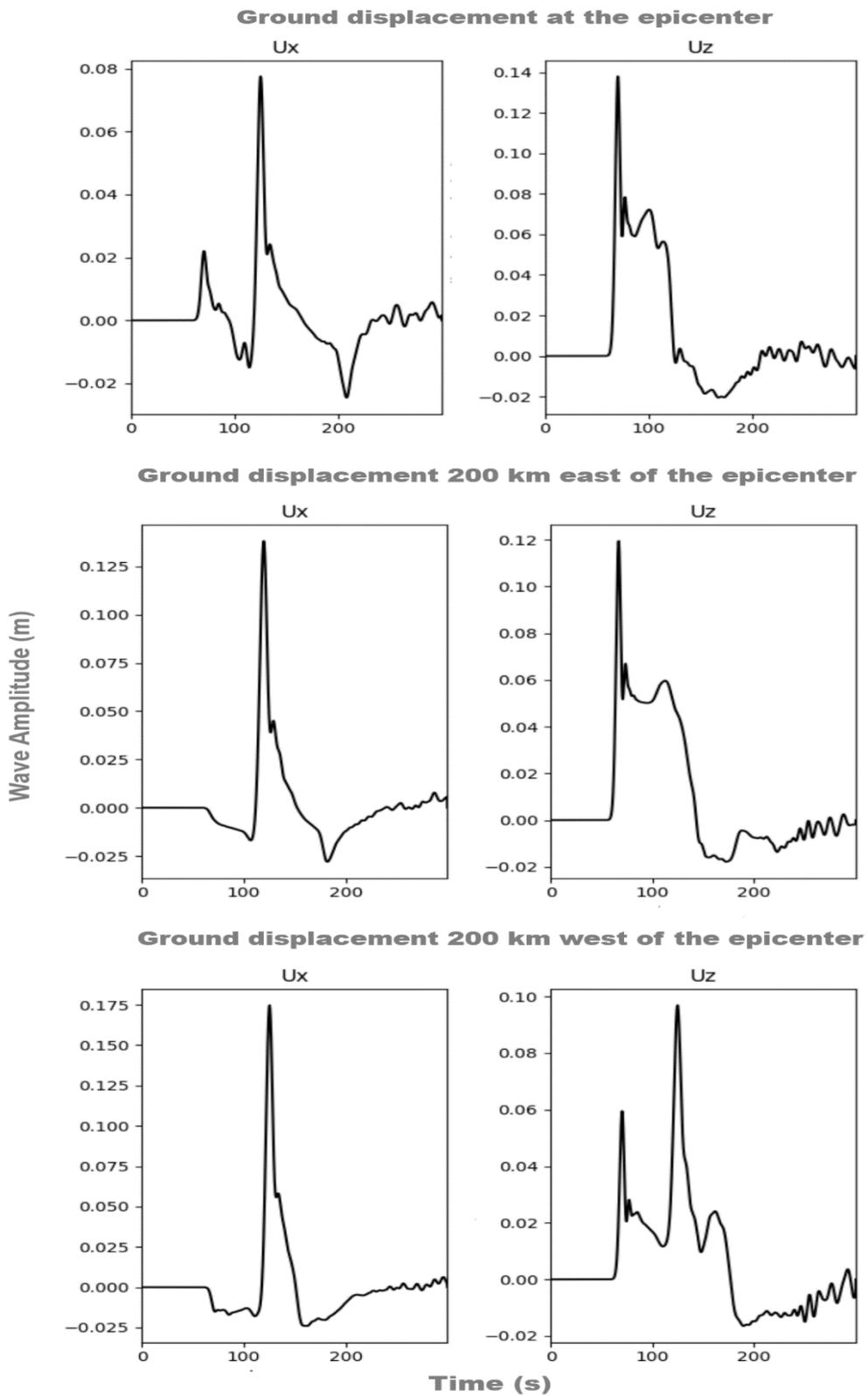

Figure 9: Theoretical seismograms of the simulated earthquake in Acre, Brazil.
Source: Author.

We can use the data obtained from the theoretical seismograms and equation (1) to estimate the magnitude $M_b$ of the simulated earthquake in Acre, Brazil. Using the epicenter function $B$ proposed by Gutenberg & Richter, (1956), and revised by Karnik et al. (1962), the magnitude $M_b$ obtained are presented in Table V.

| $u_z$ | Epicenter Function $B$ | Maximum amplitude A of the P wave ($\mu m$) | Period T (s) | Magnitude $M_b$ |
|---|---|---|---|---|
| 200 Km - West | 3.5 | 60.0000 | 53.00 | 6.55 |
| 200 Km - East | 3.5 | 120.0000 | 55.00 | 6.84 |

Table V: Theoretical results for earthquake magnitude $M_b$ of the simulated earthquake in Acre, Brazil. Source: Author

Crisostomo (2023) observed that the Acre crust showed rapid movement in a short period of time (5 years) and argues that local neotectonics played a fundamental role in this process. Santos et al. (2019) state that the frequency of seismic activity in Acre has intensified and that $37.5\%$ of events near the Tarauacá fault are of magnitude greater than $5.0\ M_b$. Theoretical studies on the intensity and frequency of earthquakes needed to explain these observed movements in Acre, over a period of 5 years, will be conducted in future work.

## CONCLUSIONS

To the best of our knowledge, there are no studies in the literature that have performed numerical simulations of intense and deep earthquakes in Brazilian territory. Specifically, earthquakes in Acre and the Tarauacá fault region have been very little studied. In this context, we argue that more research is needed on the earthquakes that have occurred in the Tarauacá fault region. Regarding these events, both experimental data and numerical simulations are lacking.

Regarding the parameters of the mathematical model, seismic analysis research in the state of Acre is still in its initial stages and requires more data and improved measurements. This research should focus on studying the properties of the surface crust region over Acre, which would allow, for example, for improved descriptions in numerical models and simulations. For our numerical studies, we need more precise data on the soil composition of the region, which will allow for better adjustments to the density and attenuation coefficient of that soil.

Regarding earthquake source modeling, we should test other possibilities. For example, instead of considering only a single source modeled by a Gaussian pulse function, we can test the use of multiple sources aligned with the Tarauacá fault, as represented in Figure 2. Would such a multiple source modeling study better simulate the deep and intense earthquakes observed in Acre, which are generally associated with the subduction of the Nazca plate under the South American plate or the relative displacement between rock blocks within the Tarauacá fault? Only new numerical simulations can answer this question.

For future research, we will test other attenuation conditions on the boundary to minimize artificial reflections, different from the conditions given in (12). Another point is to use a domain with trapezoidal geometry, more suitable for the shape of the Earth, and to adapt the numerical modeling to the generalized coordinates (Cirilo et al., 2018; Naozuka et al., 2021; Saita et al., 2021).

Finally, according to the argument of Crisóstomo (2023), is it possible that earthquakes with a magnitude around $6.0\ M_b$ explain the geomorphological transformations recently observed in the Tarauacá fault region? Theoretical studies should be conducted to verify this hypothesis.

**Acknowledgments**

We are grateful to the Coordenação de Aperfeiçoamento de Pessoal de Nível Superior (CAPES), Brazil – Finance Code 001 - for supporting the first and second authors.